# Interface Sharpening in Miscible and Isotopic Multilayers: Role of Short-Circuit Diffusion

A. Tiwari[1], M. K. Tiwari[2], M. Gupta[3], H.-C. Wille[4] and A. Gupta[1*]

[1]Amity Centre for Spintronic Materials, Amity University, U.P, Sector 125, NOIDA 201313, India

[2]Raja Ramanna Centre for Advanced Technology, Indore-452013, India

[3]UGC-DAE, Consortium for Scientific Research, Indore-452017, India

[4]Deutsches Elektronen-Synchrotron DESY, Notkestraße 85, 22607 Hamburg, Germany

[*]agupta2@amity.edu

Atomic diffusion at nanometer length scale may differ significantly from bulk diffusion, and may sometimes even exhibit counterintuitive behavior. In the present work, taking Cu/Ni as a model system, a general phenomenon is reported which results in sharpening of interfaces upon thermal annealing, even in miscible systems. Anomalous x-ray reflectivity from a Cu/Ni multilayer has been used to study the evolution of interfaces with thermal annealing. Annealing at 423 K results in sharpening of interfaces by about 38%. This is the temperature at which no asymmetry exists in the inter-diffusivities of Ni and Cu. Thus, the effect is very general in nature, and is different from the one reported in the literature, which requires a large asymmetry in the diffusivities of the two constituents [Z. Erdélyi et al., Science **306**, 1913 (2004).]. General nature of the effect is conclusively demonstrated using isotopic multilayers of $^{57}$Fe/$^{natural}$Fe, in which evolution of isotopic interfaces has been observed using nuclear resonance reflectivity. It is found that annealing at suitably low temperature (e.g. 523 K) results in sharpening of the isotopic interfaces. Since chemically it is a single Fe layer, any effect associated with concentration dependent diffusivity can be ruled out. The results can be understood in terms of fast diffusion along short-circuit paths like triple junctions, which results in an effective sharpening of the interfaces at relatively low temperatures.

## I. INTRODUCTION

Interfaces hold the key to functionality of multilayer nanostructures. Various phenomena like tunnel magnetoresistance, exchange bias, spin orbit torque, anisotropic Dzyalosinskii-Moriya interactions; depend crucially on the structure of the interfaces [1-3]. Therefore, a controlled manipulation of the interface structure is a prime requirement for tuning the functionality of multilayer nanostructures. Atomically sharp interfaces are an essential requirement for several applications, for example, i) in case of x-ray and neutron multilayer mirrors, sharp interfaces are a prerequisite in order to have high reflectivity at the Bragg peak [4,5]. ii) in a magnetic tunnel junction, magnetoresistance sensitively depends upon the sharpness of interfaces between magnetic electrodes and tunnel barrier [6], iii) sharp interfaces are also a prerequisite for interface induced perpendicular magnetic anisotropy in systems like CoFeB/MgO, Co/Pt [7-10]. This necessitates a detailed understanding of the interdiffusion at the interfaces occurring during film deposition as well as during post-deposition treatments like thermal annealing.

Typical length scales involved in such devices are in the sub nm range. Extensive studies in the literature show that interdiffusion at this length scale can be very different from that in the bulk [11-16]. Even the well-established laws of diffusion may not hold at nm length scale, as these laws are derived assuming a continuum medium, while at nm length scale, the discrete nature of atomic lattice becomes important [11]. Several counter-intuitive behavior have been observed at nm length scale; i) As a rule, grain boundary (GB) diffusion is faster than volume diffusion due to higher density of defects in GB region. However, volume diffusion in nanocrystalline $Fe_{73.5}Si_{13.5}B_9Nb_3Cu_1$ alloy is found to be faster than GB diffusion [12], ii) In nanocrystalline FeN phase, self-diffusion of Fe atoms, which are much bigger in size, is faster than N diffusion [13], iii) Strong asymmetry of interdiffusion has been observed at the two interfaces of a binary system like Fe-Si. It is found that diffusion at Fe-on-Si interface is faster than Si-on-Fe interface [14], iv) In completely miscible systems like Ni/Cu or Mo/V, with a large asymmetry in diffusivities of the two constituents, sharpening of interfaces has been observed instead of intermixing [15,16].

Atomic diffusion in Ni/Cu system has been extensively studied in the literature [10,11,14-19]. A large asymmetry exists between volume diffusivities of Cu-in-Ni and Ni-in-Cu. This asymmetry results in composition dependent diffusivities which in turn cause interfaces to sharpen on thermal annealing [15,20]. X-ray diffraction studies on coherent Cu/Ni multilayers have evidenced layer by layer diffusion of Ni at the interfaces [18]. All these studies have been done at sufficiently high temperatures at which volume diffusivity is appreciable. The associated diffusion length in this region is in the range of a few nm. On the other hand for most of the applications of such multilayers, e.g., in spintronic devices, x-ray and neutron mirrors, etc., typical layer thicknesses themselves are in the range of a few nm and even sub nm diffusion lengths can significantly alter the interface structure and hence the functional properties of multilayers. Thus, it is important to understand the atomic level processes involved in interdiffusion at relatively low temperatures which are encountered during processing and/or operation of such multilayer devices.

In the present work, we report a more general phenomenon in multilayers, which results in sharpening of interfaces even in miscible systems when there is no asymmetry in diffusivities of the two constituents. Taking Cu/Ni as a model system, it is shown that interface sharpening can occur even at temperatures well below the temperatures at which volume diffusion becomes appreciable, and the average diffusion lengths are in sub nm range. This is the temperature range in which one does not expect any asymmetry between diffusivities of Cu-in-Ni and Ni-in-Cu and therefore the mechanism proposed in the literature for the interface sharpening should not be operative. We propose that the atomic level processes involved in the observed effect are very different, and highlight the role of short-circuit diffusion along triple junctions in the temperature range of interest for processing of multilayer nanostructure devices. A triple junction is a line defect where three grains and grain boundaries meet (Fig. 1) [21]. In a polycrystalline material, intergranular region consists primarily of grain boundaries, network of which completely isolate individual crystallites from one another. Similarly, facets of grain boundaries are connected to one another at triple junction lines, which form a fully interconnected matrix without any dead-ends. With highly disordered structure, triple junctions have diffusivities that are several orders of magnitude higher [22,23] than grain boundary diffusivity, which in turn is generally significantly higher than volume diffusivity. In general, in polycrystalline materials, the volume fraction occupied by triple junctions is small and their contribution to atomic diffusion can be neglected. However, as the crystallite size decreases to nanoscale, the volume fraction occupied by them and hence their contribution to atomic diffusion becomes appreciable; Anomalously high atomic diffusion in nanocrystalline materials, even beyond what is expected from a high density of equilibrated grain boundaries, could be explained in terms of short circuit diffusion along triple junctions [23].

A more convincing evidence for the proposed mechanism of interface sharpening in the present case is obtained from the observed sharpening of the isotopic interfaces in a $^{57}$Fe/$^{natural}$Fe multilayer as observed using nuclear resonance reflectivity. Since in this case both types of the layers are chemically identical, there is no question of any concentration dependent diffusivity.

These results are also of practical importance since in general, x-ray and neutron mirrors, and magnetic multilayer for spintronic applications have typical layer thicknesses in the nm range, and their properties can get modified drastically even if the interfaces are modified at sub nm length scale.

## II. EXPERIMENTAL DETAILS

The multilayer of Cu/Ni system with nominal structure: substrate/[Cu(3.28 nm)/Ni(2.2 nm)]$_{10}$ and the isotopic multilayer of $^{57}$Fe/$^{natural}$Fe having nominal structure: substrate/[$^{57}$Fe(3.2 nm)/$^{natural}$Fe(3.2 nm)]$_{10}$ were prepared by ion beam sputtering at room temperature on Si (111) substrates. The substrate had a surface roughness of 0.56 nm and a 1.5 nm thick oxide layer at the surface, as determined using x-ray reflectivity. Base pressure in the chamber was 2×10$^{-7}$ mbar. The chamber was flushed with 5N purity Ar gas several times before starting the deposition. The base partial pressure of oxygen in the chamber as determined using a MKS residual gas analyser (model EVE-110-001) was 1 x 10$^{-9}$ mbar. For the deposition of multilayer structure, the targets of Cu (99.99% purity) and Ni (99.99% purity) were alternatively sputtered by Ar ion beam of 1000 eV from a 3 cm broad beam Kaufman-type ion source. For preparation of isotopic multilayers targets of $^{natural}$Fe (99.99% purity) and $^{57}$Fe (99.95% purity and 95% enrichment) were used. In the preparation of both the multilayer structures, the flow of Ar gas in the deposition chamber was controlled by a mass flow controller (MKC-MFC 1179A) at 5 SCCM (standard cubic centimeters per minute). Oxygen partial pressure in the chamber during deposition remained in the range of 10$^{-9}$ mbar.

In order to achieve variation in grain size in the isotopic multilayer, another isotopic multilayer having structure substrate/$^{natural}$Fe(50 nm)/[$^{57}$Fe(1.8 nm)/$^{natural}$Fe(3.8 nm)]$_{10}$ was prepared at an elevated substrate temperature of 573K, using direct current magnetron sputtering at a constant power of 50 W. The base pressure in the chamber was 1.7×10$^{-7}$ mbar, while the pressure during deposition was 3.4×10$^{-3}$ mbar. The flow of Ar gas in the deposition chamber was kept at 50 SCCM.

X-ray diffraction measurements were done using Bruker D5000 diffractometer in θ-2θ geometry. Anomalous x-ray reflectivity measurements have been done at beamline BL-16 of Indus-2 in order to study the interdiffusion in Cu/Ni multilayers. This technique is capable of measuring diffusion length down to accuracy of 0.1 nm. Since the electron density contrast between Cu and Ni is negligible, x-ray reflectivity measurements have been done across the K absorption edges of Ni, so as to increase the x-ray scattering contrast. It may be noted that the refractive index of Ni would vary rapidly with energy around the absorption edge therefore even a small deviation in the x-ray energy from the designated value can cause significant variation in the refractive index. In order to avoid possible artifact in the fitting of the reflectivity pattern due to some deviation in the x-ray energy from the designated value, measurements were done at five different x-ray energies across the absorption edge of Ni, namely, 8290 eV, 8315 eV, 8340 eV, 8365 eV and 8390 eV. Simultaneous fitting of all five reflectivity patterns using the same structural parameters for the multilayer, provides much more reliable and unambiguous fitting of the data. The refractive indices of Cu and Ni at various energies were taken from CXRO data tables [24]. The fitting of reflectivity data was done using Parratt's formalism [25].

Variation in the roughness of isotopic interfaces in $^{57}$Fe/$^{natural}$Fe multilayer was studied using nuclear resonance reflectivity (NRR) measurements done at the Dynamics Beamline P01 at PETRA-III, DESY, Hamburg. To perform NRR measurements, the energy of the radiation was kept at 14.4 KeV (Mössbauer resonance of $^{57}$Fe) in 40 bunch mode with the bunch separation of 192 ns. The delayed events, resulting from the nuclear forward scattering, were counted by a fast avalanche photodiode (APD) detector.

The isotopic multilayer sample was cut in to 10x10 mm pieces and each piece was annealed at a particular temperature for 1h. Cu/Ni multilayer samples were isochronally annealed at different temperatures in a tubular furnace having a uniform temperature zone of 5 cm. The pressure in the tube was $1\times10^{-6}$ mbar and the temperature was maintained within ±1°C of the set temperature during annealing.

## III. RESULTS AND DISCUSSIONS

### (a) Cu/Ni multilayer

Figure 2 gives the X-ray diffraction pattern of as-prepared Cu/Ni multilayer taken in θ-2θ geometry, with scattering vector normal to the film surface. One observes a dominant peak around 2θ = 44°. Both Ni as well as Cu have fcc structure with their (111) peak at 45° and 43° respectively. The observed peak is in between these angular positions. The crystallite size as obtained from the width of diffraction peak using Scherrer formula comes out to be 9.7 nm, which is about 4 times the thickness of individual Ni or Cu layer. This means that a crystallite, on the average, extends over 4 adjacent layers, suggesting a partial coherency between layers of Ni and Cu. Appearance of weak sidebands around (111) peak with separation from the main peak corresponding to the periodicity of the multilayer is also suggestive of partial coherence between Ni and Cu grains. At such a small grain size of 9.7 nm, even the grain boundaries and triple junction defects would occupy a significant volume fraction of the film. Taking typical width of the grain boundaries to be 1 nm, the volume fraction occupied by grain boundaries and triple junctions can be calculated to be 25% and 3% respectively [26].

Figure 3a gives the x-ray reflectivity of pristine Cu/Ni multilayer at five different energies across the absorption edge of Ni. Energies of 8290 eV and 8315 eV are below the absorption edge of Ni and 8340 eV, 8365 eV, 8390 eV are above the absorption edge of Ni. Variation in the refractive indices and hence in the scattering contrast between Ni and Cu occurs as the energy of the x-rays varies, the contrast being maximum at 8340 eV, which is just above the absorption edge of Ni. The height of the Bragg peak exhibits a systematic variation with energy because of the variation in the scattering contrast. All five reflectivity patterns were fitted simultaneously by taking the same set of values of parameters like thicknesses and electron densities of Ni and Cu layers and their interface roughnesses.

Figure 3b gives the reflectivity pattern of Cu/Ni multilayer isochronally annealed at various temperatures for one hour each. Results are presented for x-ray energy of 8340 eV. The continuous curves in the figure represent the best fit to the experimental data. The results of fitting are summarized in table 1. One can see that the roughness of Cu-on-Ni is significantly higher than that of Ni-on-Cu. This result is in general agreement with the literature [16]. With thermal annealing, the only parameter which exhibits significant variation is the interface roughness. One expects that with thermal annealing as the interdiffusion between Cu and Ni layers occurs, the interface roughnesses would increase and the height of the Bragg peak should come down. One can calculate the diffusion length from the relative decrease in the height of the first Bragg peak using the relation [27, 28]:

$$\ln\left[\frac{I(t)}{I(0)}\right] = \frac{-8\pi^2 D(T) t}{d^2} \quad , \qquad (1)$$

where $I(0)$ and $I(t)$ are the intensities of the Bragg peak before and after annealing at temperature $T$ for a time $t$. $D(T)$ is the diffusivity at temperature $T$ and $d$ is the bilayer period.

Surprisingly after annealing at 423 K for 1 h, the height of the Bragg peak exhibits an enhancement relative to that in the as-deposited multilayer, suggesting suppression in the interface roughnesses. Beyond 423 K the Bragg peak exhibits a systematic reduction in the height, as expected on the basis of interdiffusion. Figure 4 gives the diffusivity at 473 K and 523 K as calculated using Eq. 1. For comparison the volume, grain boundary and triple junction diffusivities in Cu/Ni system taken from literature [29,30] are also shown. One can see that calculated diffusivities lie roughly along the line for grain boundary diffusion.

On another piece of the multilayer specimen, reflectivity measurements were done as a function of the annealing time at 423 K. Bilayer period of this piece happened to be about 4% lower than that of the earlier one. Figure 5 gives the reflectivity pattern of the specimen after annealing at 423 K for 30 min, at which the maximum increase in the height of the Bragg peak was observed. The reflectivity at the first order Bragg peak increases from 0.7% in as-prepared multilayer to 2% after 30 min annealing, and corresponding decrease in the interface roughness is from 1.27 nm to 0.65 nm on Ni-on-Cu interface and from 1.6 nm to 1.1 nm on Cu-on-Ni interface respectively. Annealing for longer time results in decrease in the reflectivity. Diffusivity at 423 K was calculated by taking the decrease in the height of the Bragg peak after 60 min annealing relative to that after 30 min and is also shown in Fig. 4. One may note that the calculated diffusivity comes out to be orders of magnitude higher than the extrapolated value of grain boundary diffusion at this temperature.

It may be noted that both experimental as well as theoretical studies have shown that in miscible systems having a large asymmetry in the diffusivities of the two constituents, diffusion can result in sharpening of the interfaces. In Mo/V system, annealing up to 973 K resulted in sharpening of interfaces from 1.7 nm and 1.4 nm respectively to 0.78 nm [15]. In Ni/Cu system also, interface sharpening has been observed upon annealing at 773 K [16]. The observed effect has been attributed to large asymmetry in the diffusivities of the two constituents, which leads to a concentration dependent diffusivity in these systems. With increasing Ni concentration the diffusivity of Cu decreases while that of Ni increases. Also, studies have shown that while the volume diffusivities of Cu and Ni are very different, no such asymmetry exists in grain boundary or triple junction diffusivities of the two constituents [19]. Thus the observed interface sharpening in the literature is a consequence of a large asymmetry in the volume diffusivities of the two constituents. Indeed the temperature regime in which interface sharpening has been observed in the literature is such that volume diffusion is appreciable. In contrast, in the temperature range studied in the present work, volume diffusivity is negligible. At 423 K the extrapolated values of grain boundary and triple junction diffusivities are $2\times10^{-23}$ m$^2$ s$^{-1}$ and $4\times10^{-20}$ m$^2$ s$^{-1}$ respectively [29]. Therefore, at this temperature, only triple junction diffusion is expected to be appreciable. Thus, asymmetry in the diffusivities of the two constituents cannot be the cause of the observed interface sharpening in the present study.

We propose a mechanism for the observed sharpening of the interfaces at 423 K, in which fast diffusion along the short-circuit paths like triple junctions plays a dominant role. It may be noted that in as-deposited multilayer certain amount of intermixing exists at the interfaces of the two constituents as a consequence of the finite kinetic energy of the impinging atoms which make some diffusive jumps before settling down to a lattice site. The extent of intermixing will depend on the activation energy involved in a diffusive jump. The smaller the activation energy the larger the intermixing. Since the activation energy for diffusive jumps along a triple junction is orders of magnitude lower than that for grain boundary or volume diffusion, the intermixing between Cu and Ni across a grain or a grain boundary should be much smaller than that across a triple junction line which intersects the interface between two Ni and Cu layers. The situation is depicted schematically in Fig. 6a. The interface roughness has two contributions, the correlated part $\sigma_c$ which propagates from the substrate itself, and the uncorrelated part $\sigma_u$ [4]. The uncorrelated part in turn consists of topological term due to stochastic nature of the thin film deposition ($\sigma_{uT}$) and the interdiffusion term ($\sigma_{uD}$). Thus the total roughness can be written as:

$$\sigma^2_T = \sigma^2_c + \sigma^2_u \quad , \tag{2}$$

where,

$$\sigma^2_u = \sigma^2_{uT} + \sigma^2_{uD}. \tag{3}$$

Annealing at 423 K causes intermixing between Cu and Ni layers through fast interdiffusion along those triple junction lines which intersect the interfaces. Due to orders of magnitude higher triple-junction diffusivity as compared to gran-boundary or volume diffusivities at this temperature, one expects type C diffusion. Once composition along such triple junctions homogenizes, contribution of triple junctions to $\sigma^2_{uD}$ will disappear (Fig. 6b). Instead, as a result a small decrease in the scattering contrast between Cu and Ni layers will appear. The net result of this is an effective sharpening of the interfaces. Continued annealing at 423 K will result in further composition variation along the triple junctions which do not intersect the interfaces. This should result in a decrease in the scattering contrast between adjacent layers, resulting in a decrease in the height of the Bragg peak. Annealing at still higher temperatures of 473 K and 523 K at which grain boundary diffusion becomes appreciable, interface roughness will increase due to partial intermixing along grain boundaries. Indeed the diffusivities calculated from the decrease in the height of the first order Bragg peak at 473 K and 523 K lies close to the extrapolated value of grain boundary diffusion as determined in the literature [29].

It may be noted that the diffusivities as obtained in the present case show a small decrease with increasing temperature. This happens because of the transient nature of diffusion in this temperature-time regime and the fact that a single sample is isochronally annealed at successively higher temperatures and the calculated diffusivities are the average diffusivities over the annealing period. As discussed above, at 423K observed diffusivity is intermediate between triple junction and grainboundary diffusivities. At 473K the diffusion along triple junctions is already saturated and further diffusion occurs along the grain boundaries. During annealing at 523K, initially diffusion occurs along grain boundaries and once the grain boundaries are saturated, further diffusion occurs through grains. Therefore the average diffusivity at this temperature is intermediate between that for grain boundary and volume diffusivities.

**(b) $^{57}$Fe / $^{natural}$Fe multilayer**

In order to conclusively exclude the possibility of some concentration dependent diffusivity being the cause of interface sharpening in Cu/Ni multilayer, effect of thermal annealing on the isotopic interfaces in a multilayer consisting of nanocrystalline $^{57}$Fe/$^{natural}$Fe was studied using nuclear resonance reflectivity (NRR). Since chemically the isotopic multilayer is a single layer of Fe, there is no question of any concentration dependent diffusivity in the system. When the energy of the x-rays is tuned to the Mössbauer transition of $^{57}$Fe nuclei, nuclear resonance scattering from $^{57}$Fe causes a strong scattering contrast between $^{57}$Fe and $^{natural}$Fe. This results in Bragg peaks in the nuclear resonance reflectivity corresponding to the isotopic periodicity in the multilayer [27]. The height of a Bragg Peak is proportional to the roughness of the isotopic interfaces and has been used in the literature to study self-diffusion of Fe [27,31,32]. Figure 7a gives the nuclear resonance reflectivity of an isotopic multilayer having structure [$^{57}$Fe (3.2 nm)/$^{natural}$Fe (3.2 nm)]$_{10}$ as a function of annealing at different temperatures for 1 h each. The reflectivity patterns have been fitted using the software REFTIM [33] and the simulated results are shown as continuous curves. One expects an error function concentration profile at the interface, however, for simplicity a linear concentration profile was assumed. The calculated concentration profile of $^{57}$Fe is shown in Fig. 7b. One can see that annealing at 523 K results in an increase in the reflectivity at the first Bragg peak, and a sharpening of the isotopic interface as seen from Fig. 7b. In the isotopic multilayer, since both the constituents are chemically identical, the mechanism of concentration dependent diffusivities proposed in the earlier studies can be ruled out. In this system, the mechanism proposed in the earlier section is the only possibility. Negative diffusion length at 523 K as calculated from the height of the Bragg peak comes out to be -0.61 nm. The calculated value of diffusivity at 573 K and 623 K comes out to be $8.9 \times 10^{-23}$ m$^2$ s$^{-1}$ and $1.37 \times 10^{-22}$ m$^2$ s$^{-1}$ respectively. The present values lie in between those for volume [34] and grain boundary diffusivities [35,36] and are close to the values for self-diffusion of Fe in nanocrystalline film as obtained in some earlier study [37].

It may be noted that grain boundary diffusivities reported in the literature are for polycrystalline Fe. The same in nanocrystalline film with a higher density of grain boundaries is expected to be higher. Thus, it may be noted that the grain boundary diffusion even at 523 K is not negligible and should cause an increase in the interface roughness. However, the effective sharpening of the interfaces associated with homogenization of isotopic composition along the triple junctions dominates, causing an overall decrease in the interface roughness. In order to further ascertain this point, another isotopic multilayer having structure substrate/ $^{natural}$Fe(50 nm)/[$^{57}$Fe(1.8 nm)/$^{natural}$Fe(3.8 nm)]$_{10}$ was prepared by keeping the substrate temperature at 573 K. Nuclear resonance reflectivity of the as prepared as well as 573 K annealed multilayer is shown in Fig. 8. The negative diffusion length as calculated from the height of the Bragg peak comes out to be -0.55 nm, which is less than the value obtained after annealing of room temperature multilayer at 523 K. The result can be understood by considering that grain boundary diffusion at 573 K is higher than that at 523 K, and therefore the roughening effect of grain boundary diffusion is higher in this case.

## IV. CONCLUSIONS

In conclusion, the evolution of interfaces in Cu/Ni multilayer with thermal annealing has been studied using anomalous x-ray reflectivity. It is found that at relatively low temperature of 423K, at which both volume as well as grain boundary diffusion are low, thermal annealing results in sharpening of the interfaces, causing interface roughnesses to decrease by about 38%. Annealing at higher temperatures in the range 473K to 523K, results in interdiffusion and the calculated diffusivities agree with extrapolated values of grain-boundary diffusivities from the literature. It is suggested that fast diffusion occurring along the short-circuit paths like triple junctions causes an effective decrease in the interface roughnesses. Sharpening of the interfaces is expected to occur in the temperature regime in which diffusion along short-circuit paths is fast enough, while the normal diffusion is significantly lower. The observed effect is general in nature and different from that observed in systems with a large asymmetry in the diffusivities of the two constituents [15,16]. The general nature of the effect is conclusively demonstrated by studying the evolution of isotopic interfaces in $^{57}$Fe/$^{natural}$Fe multilayers using nuclear resonance reflectivity. Sharpening of isotopic interfaces is observed upon annealing at 523K. In the isotopic multilayer, since both the constituents are chemically identical, the mechanism of concentration dependent diffusivities proposed in the earlier studies can be ruled out.

**ACKNOWLEDGEMENTS**

This work was partially supported by Science and Engineering Research Board through project no. SB/S2/CMP-007/2013. Support for the experiment at PETRA-III, Germany, was provided by the Department of Science and Technology, Govt. of India through Jawaharlal Nehru Centre for Advanced Scientific Research. Fitting of NRR data was done using the program REFTIM, kindly provided by Prof. Marina A. Andreeva. Thanks are due to Ms. Jagrati Dwivedi and Mr. Layant Behra for their help in deposition of multilayer structures and for XRD measurements. Help of Mr. Ajay Khooha during measurements at BL-16 of Indus-2 is thankfully acknowledged.

FIG. 1 Schematic representation of grains, grain boundaries and triple junction.

FIG. 2. XRD of Cu/Ni multilayer in the as-prepared state.

FIG. 3. (a) X-ray reflectivity of pristine Cu/Ni multilayer taken at different x-ray energies across the absorption edge of Ni. The continuous curves represent the best fit to the data. (b) X-ray reflectivity of Cu/Ni multilayer as a function of isochronal annealing at different temperatures for 1 h each. The continuous curves represent the best fit to the data. Various reflectivity curves are shifted vertically with respect to each other for the clarity of presentation.

FIG. 4. Coefficients for interdiffusion in Cu/Ni system for volume, grain boundary and triple junction diffusion taken from the literature [29, 30]. The dashed lines represent extrapolation of literature data to lower temperatures. Stars represent the average diffusivities at different temperatures as obtained in the present work. Typical error bars in the calculated values are about 10%.

FIG. 5. X-ray reflectivity of Cu/Ni multilayer annealed at 423 K for 0.5 h. Inset compares the reflectivity at the Bragg peak of samples annealed at different temperatures.

FIG. 6. Schematic of the structure of the multilayer before and after annealing, exhibiting different interdiffusion across Cu/Ni interfaces inside crystallites and in triple junction regions. Open circles represent Ni atoms, while filled circles represent Cu atoms. Typical triple-junction-line extending across several Ni and Cu layers, and those confined within a single Ni or Cu layer are indicated by arrows. Due to limitation with the projection on 2-D plane, grain boundary regions could not be depicted. It may be noted that there is a partial epitaxy between Cu and Ni layers with the average crystallite size of 9.7 nm, thus a typical crystallite extends over several layers. Difusion profile across the interfaces is also shown.

FIG. 7. (a) Nuclear resonance reflectivity of isotopic multilayer substrate/[$^{57}$Fe(3.2 nm)/$^{natural}$Fe(3.2 nm)]$_{10}$ after annealing at different temperatures for 1h each. The continuous curve represents the best fit to the experimental data, assuming a linear concentration profile across the isotopic interface. (b) The corresponding concentration profiles as obtained from the best fit of the data.

FIG. 8. Nuclear resonance reflectivity of isotopic multilayer substrate/$^{natural}$Fe(50 nm)/[$^{57}$Fe(1.8 nm)/$^{natural}$Fe(3.8 nm)]$_{10}$ prepared at 573 K in pristine form and after annealing at 573 K for 1h.

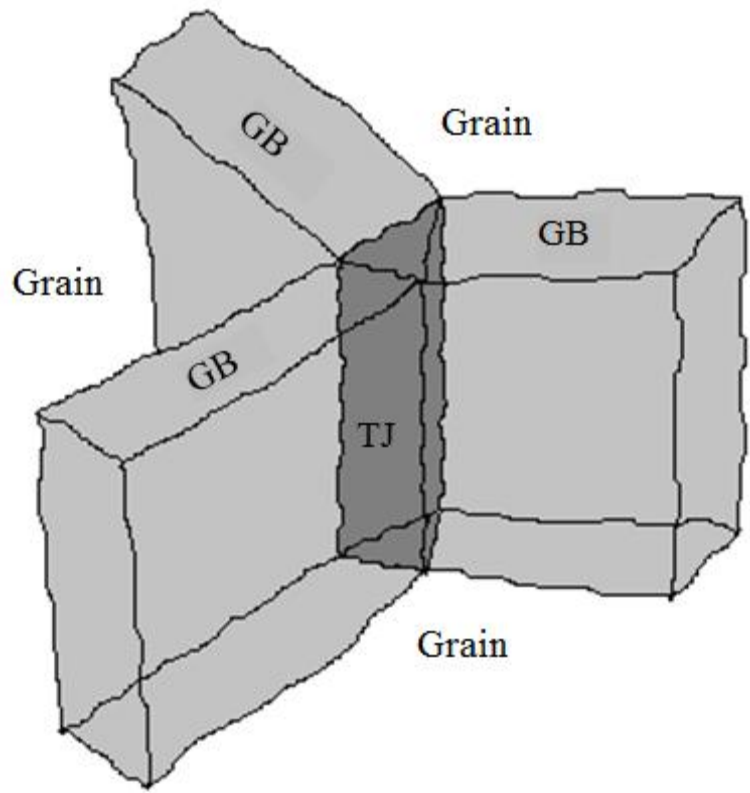

FIG. 1.

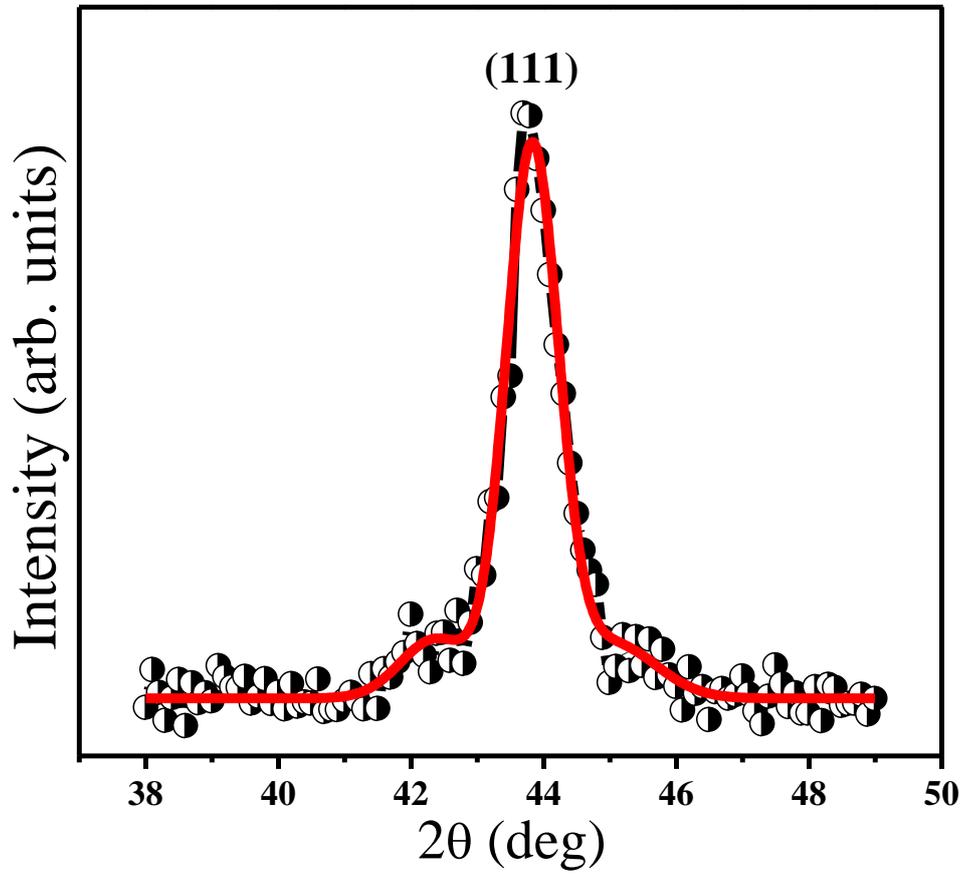

FIG. 2.

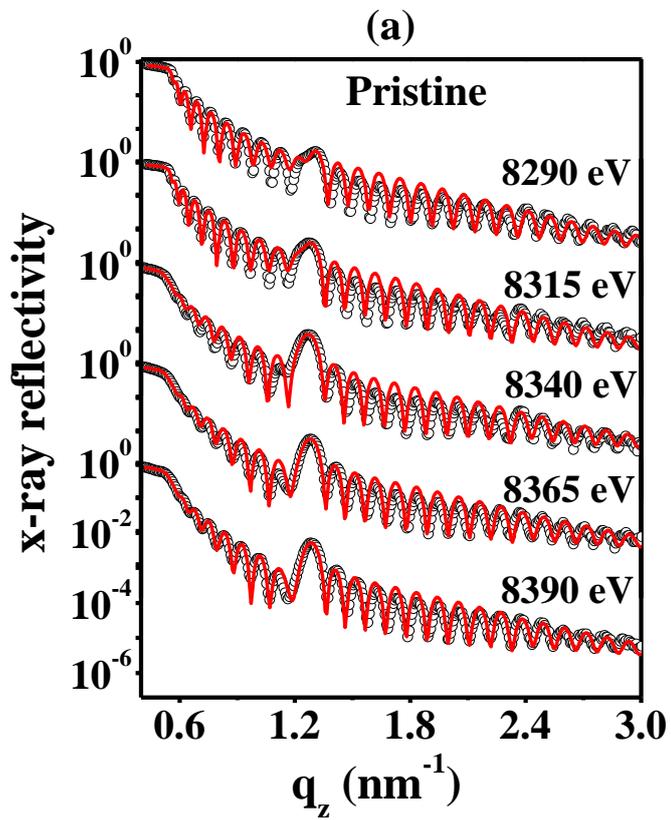 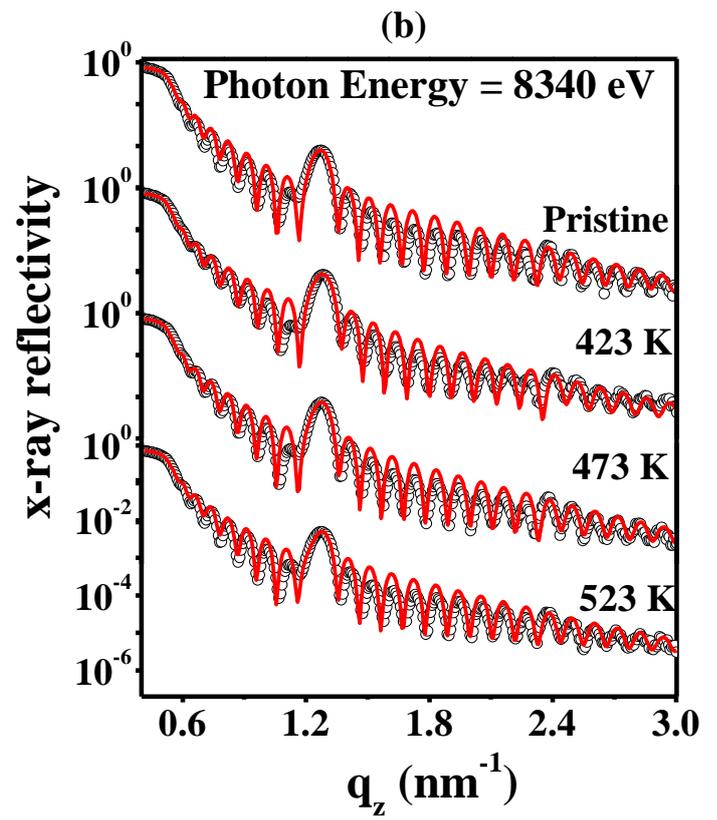

FIG. 3.

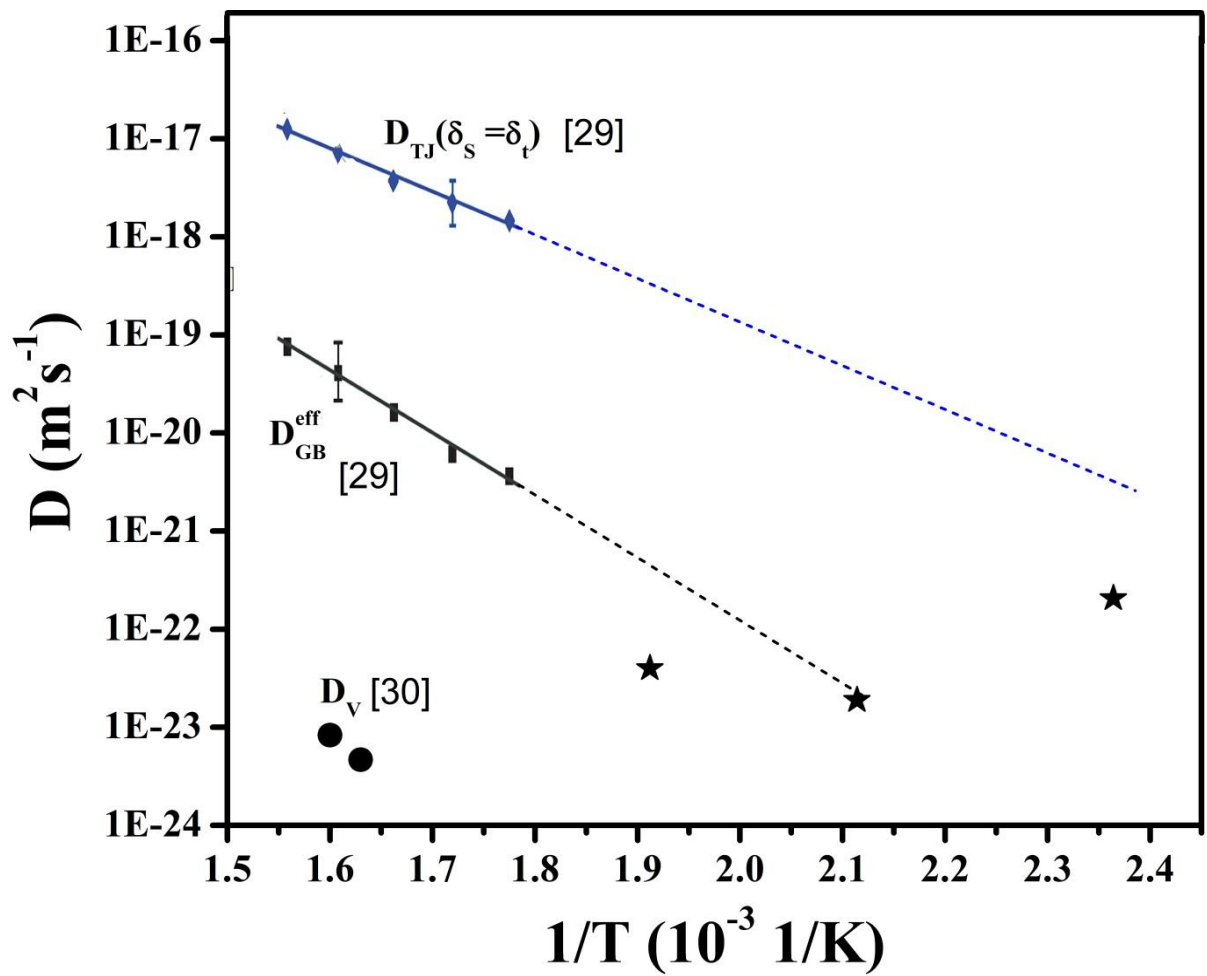

FIG. 4.

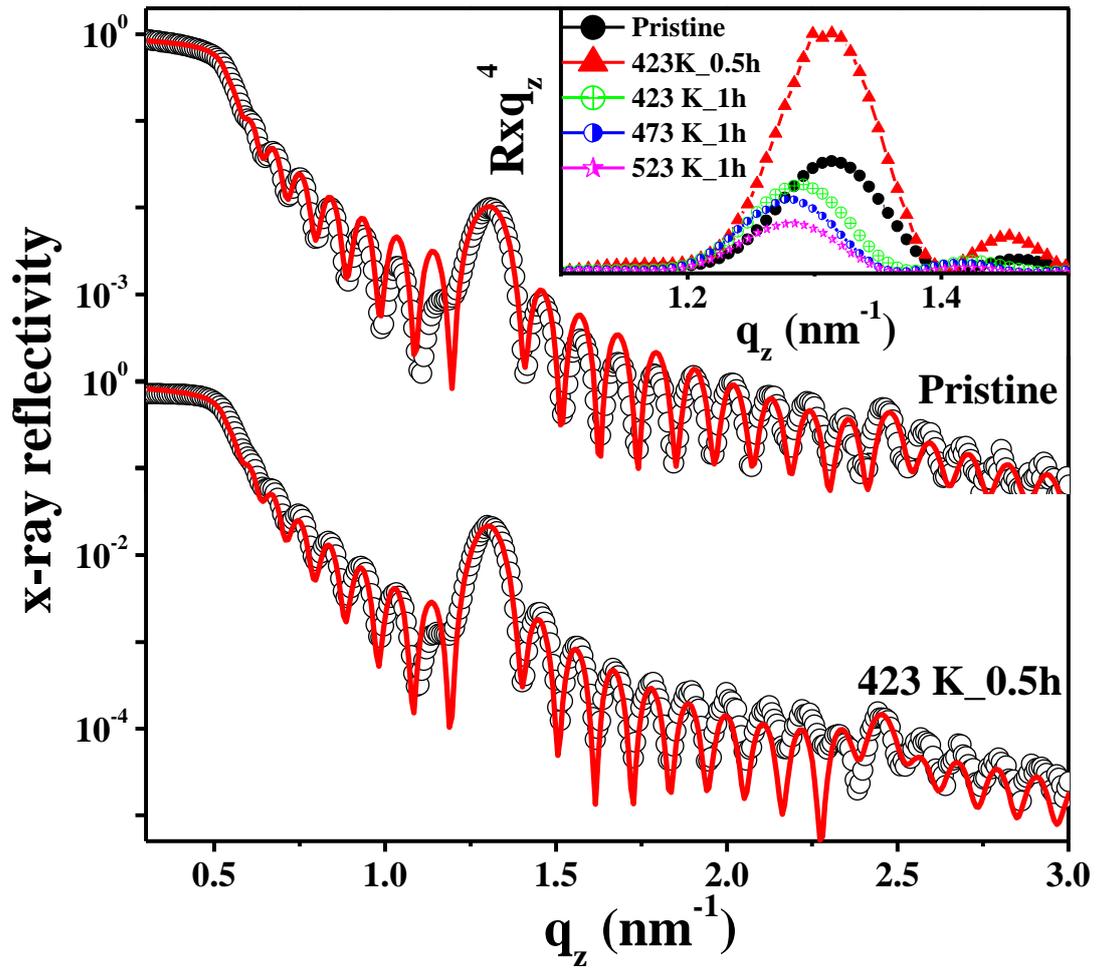

FIG. 5.

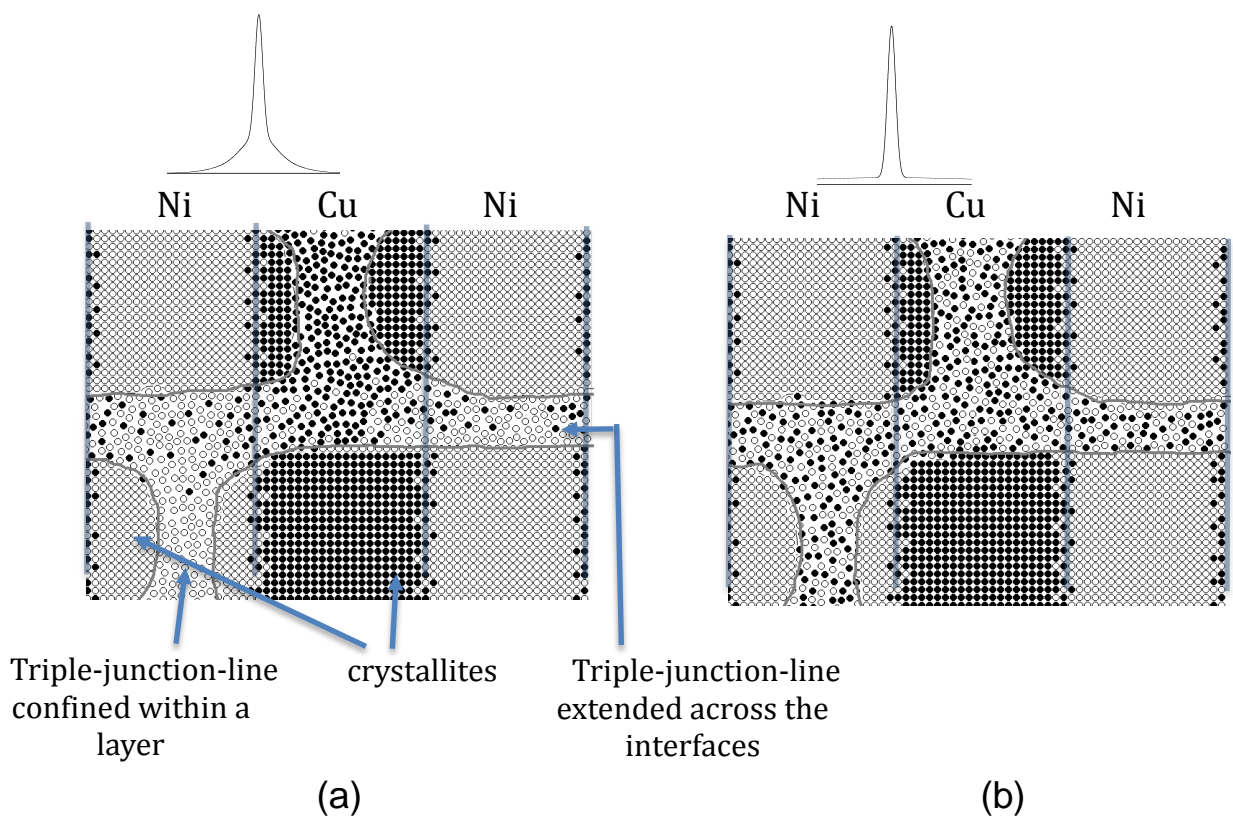

FIG. 6.

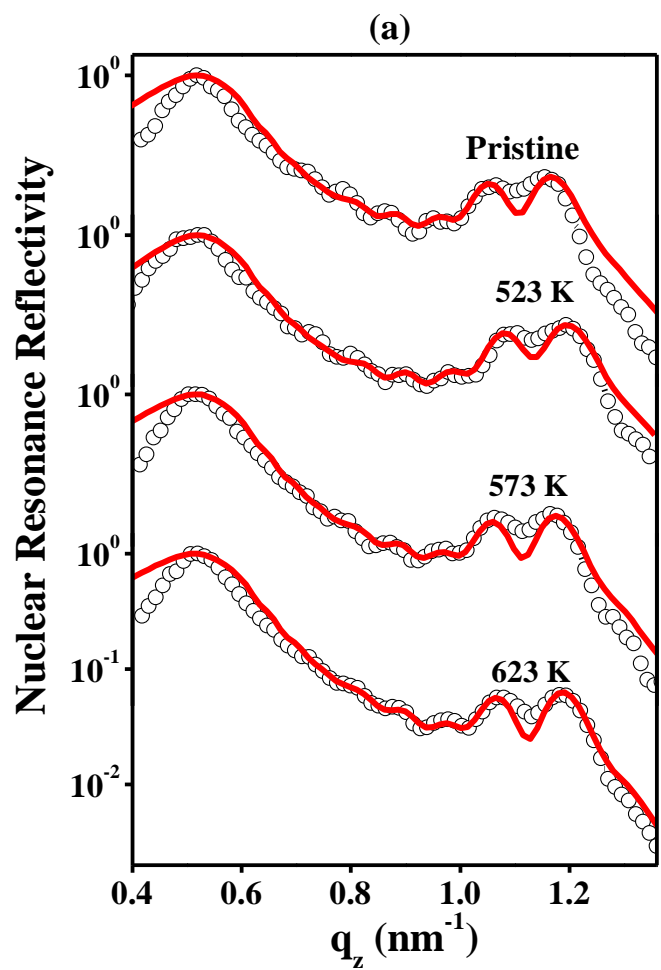 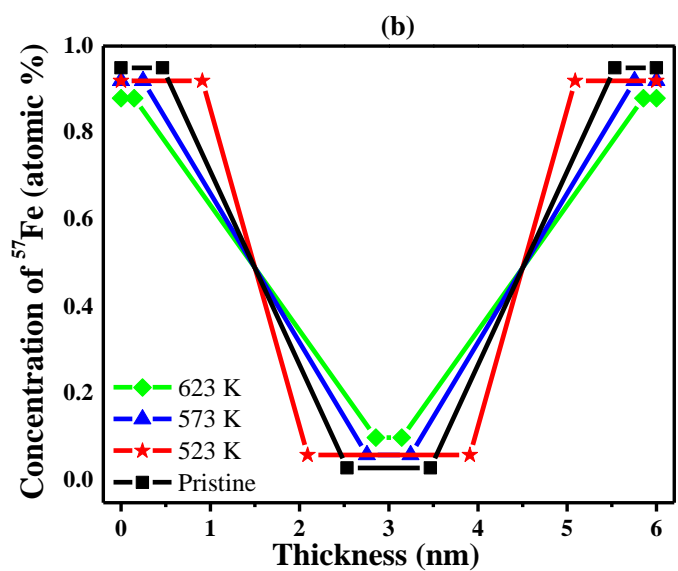

FIG. 7.

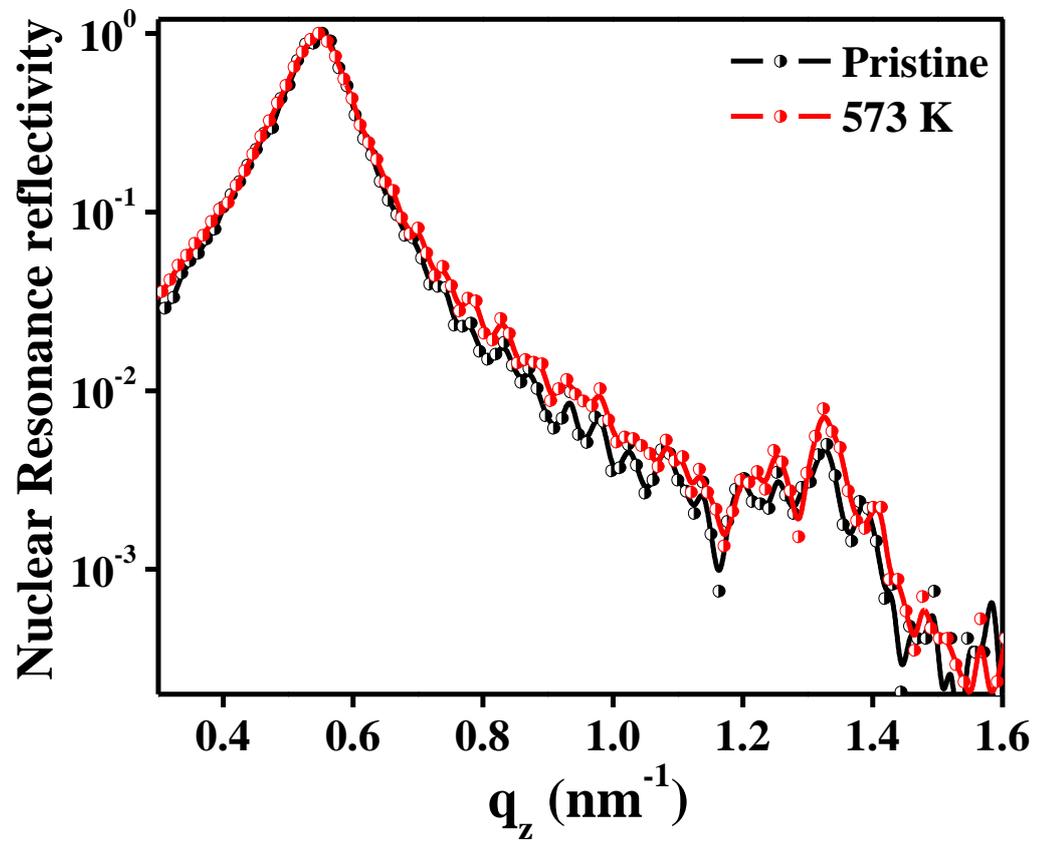

FIG. 8.

TABLE I. Results from the fitting of anomalous x-ray reflectivity of Cu/Ni multilayer as a function of isochronal annealing at different temperatures for 1 h each. The results of the sample annealing at 423 K at 0.5 h are also included.

| Temperature (K) | Cu | | | Ni | | | Chi$^2$ |
|---|---|---|---|---|---|---|---|
| | Thickness (nm) | Roughness (nm) | Scattering length density (nm$^{-2}$) | Thickness (nm) | Roughness (nm) | Scattering length density (nm$^{-2}$) | |
| **RT** | 3.27±0.05 | 1.27±0.05 | 6.38×10$^{-4}$ | 2.2±0.05 | 1.6±0.05 | 4.4×10$^{-4}$ | 1.215 |
| **423 (0.5 h)** | 3.2±0.05 | 0.65±0.05 | 6.28×10$^{-4}$ | 2.1±0.05 | 1.1±0.05 | 4.5×10$^{-4}$ | 0.216 |
| **423** | 3.24±0.05 | 1.08±0.05 | 6.28×10$^{-4}$ | 2.18±0.05 | 1.45±0.05 | 4.5×10$^{-4}$ | 0.726 |
| **473** | 3.26±0.05 | 1.18±0.05 | 6.28×10$^{-4}$ | 2.2±0.05 | 1.55±0.05 | 4.5×10$^{-4}$ | 0.965 |
| **523** | 3.26±0.05 | 1.33±0.05 | 6.28×10$^{-4}$ | 2.2±0.05 | 1.68±0.05 | 4.5×10$^{-4}$ | 1.353 |